\documentclass[aps,preprint,showpacs,floatfix]{revtex4}
\usepackage{epsfig}
\usepackage{graphicx}
\usepackage[normalem]{ulem}
\RequirePackage{color}

\definecolor{MyDarkGreen}{rgb}{0.02,0.60,0.06}

\begin{document}

\title{Universality and exact finite-size corrections for   spanning trees on  cobweb and fan networks}

\author{Nickolay Izmailian}
\email{izmail@yerphi.am; ab5223@coventry.ac.uk}
\affiliation{A. Alikhanyan National Laboratory (Yerevan Physics Institute), Alikhanian Brothers 2, 375036 Yerevan, Armenia}
\affiliation{Bogoliubov Laboratory of Theoretical Physics, Joint Institute for Nuclear Research, 141980 Dubna, Russian Federation}
\author{Ralph Kenna}
\email{r.kenna@coventry.ac.uk} \affiliation{Statistical Physics Group, Centre for Fluid and Complex Systems, Coventry University, Coventry CV1 5FB, UK}
\date{\today}

\begin{abstract}
The concept of universality is a cornerstone of theories of critical phenomena. 
It is very well understood in most systems, especially in the thermodynamic limit.
 {Finite-size systems present additional challenges. 
Even in low dimensions,} universality of the edge and corner contributions to free energies and response functions is less investigated and less well understood. 
In particular, the question arises how universality is maintained in correction-to-scaling in systems of the same universality class but with very different corner geometries. 
Two-dimensional geometries deliver the simplest such examples that can be constructed with and without corners.  

To investigate how the presence and absence of corners  {manifest universality,} we analyze the spanning tree generating function on two different finite systems, namely the cobweb and fan networks. 
The corner free energies of these configurations have stimulated significant interest precisely because of expectations regarding their universal properties and we address how this can be delivered given that the finite-size cobweb has no corners while the fan has four.

To answer, we appeal to the Ivashkevich-Izmailian-Hu approach which unifies the generating functions of distinct networks in terms of a single partition function with twisted boundary conditions. 
This unified approach shows that the contributions to the individual corner free energies of the fan network sum to zero so that it precisely matches that of the web. 
It therefore also matches conformal theory (in which the central charge is found to be c = - 2) and finite-size scaling predictions. 

Correspondence in each case with results established by alternative means for both networks verifies the soundness of the Ivashkevich-Izmailian-Hu algorithm. 
Its broad range of usefulness is demonstrated by its application to hitherto unsolved problems – namely the exact asymptotic expansions of the logarithms of the generating functions and the conformal partition functions for fan and cobweb geometries. 
We also investigate strip geometries, again confirming the predictions of conformal field theory.

Thus the resolution of a universality puzzle demonstrates the power of the algorithm and opens up new applications in the future.

\end{abstract}


\maketitle

\section{Introduction}
\label{Introduction}
Spanning trees are of  fundamental and practical importance for {connected graphs} as they represent the most efficient manner in which every node in the structure can be {linked}. 
Their enumeration is a {famous challenge in} combinatorial graph theory {originally} considered by Kirchhoff  in the context of electrical networks \cite{Kirchhoff}. 
Besides being of  interest in mathematics as a fundamental challenge \cite{Burton,Lyons}, enumeration of such trees remains of importance to other disciplines such as physics, engineering and computing \cite{Wu1977,Jaya}. 
Diverse contexts include standard \cite{Noh} and loop-erased random walks \cite{Dhar}, and spanning tree numbers can be mapped to the partition function of the $q$-state Potts model \cite{Wu1982} in the limit of $q$ approaching zero. 
{Another limit is $v \rightarrow 0$, where $v=e^J -1$, and $J$ is the coupling constant, with different fixed values 
of $v/q$ delivering the generating function for spanning trees, or spanning forests, etc.
\cite{Jesperandfriends}.}
{Another close  relationship is  to} the sandpile model \cite{Bak}.
Because of this diversity of applications~\cite{WuChao}, mathematical investigations of spanning trees continue apace, building on the considerable progress already achieved.
The exact numbers of spanning trees have been determined for regular lattices \cite{Wu1977,Tzeng,Wu2000}, networks \cite{Izmailian2014, Izmailian2015}  and Sierpinski gaskets \cite{Chang}, and the well-known bijection between close-packed dimer coverings and spanning tree conﬁgurations on two related lattices \cite{Temperley} add to their attractiveness as a challenging realm of study.

Systems which are confined by various boundary conditions when finite in extent have the same per-site properties in the bulk limit - e.g., free energy, internal energy, and specific heat.
However, boundary  characteristics are manifest in the correction terms of finite systems.
The study of finite-size scaling and corrections to scaling {in this context} was {instigated} {over 40 years ago} ago by Fisher and Barber \cite{barber} and continues to attract a great deal of attention both for fundamental and applied purposes  (see, e.g., \cite{privmanbook,hu} for well-known reviews
{and \cite{ref3a,ref3b} for corner and boundary effects in Ising and Potts models}).
Theoretical interest includes extrapolating finite or partially-finite systems to determine critical and non-critical properties of their infinite  counterparts.
More practical interests in finite-size effects stem from  recent progress in fine processing technologies {for}  nanoscale materials with novel shapes which has recently been enabled \cite{nano1,nano2,nano3}.
To understand scaling and corrections terms,  exact results are of prime interest because only in these cases can the analysis deliver results without the vagaries of numerical errors.

In 2002 Ivashkevich, Izmailian, and Hu \cite{Izmailian2002} developed a method which delivers exact finite-size corrections for various functions of fundamental importance - e.g., partition functions and their derivatives.
The method was applied to iconic models of statistical physics such as the Ising, dimer, and Gaussian models and the results demonstrated that the partition function of each model can be written as a generic partition function with twisted-boundary conditions, viz $Z_{\alpha,\beta}$ with $(\alpha,\beta) = (1/2,0),(0,1/2)$, and $(1/2,1/2)$.
Building  upon this approach, computations by Izmailian, Oganesyan, and Hu \cite{Izmailian2003} delivered finite-size corrections to the free energy of the square-lattice dimer model under five distinct sets of boundary conditions (namely free, cylindrical, and toroidal boundaries as well as the M$\ddot{o}$bius strip and the Klein bottle).
{The} dependence of finite-size corrections on the aspect ratio {was found to be} sensitive to boundary conditions as well as to the parity of the number of lattice sites along the lattice axes.
In 2014  Izmailian et al. \cite{Izmailian2014a}  found that {in the case of a rectangular $(2M - 1) \times (2N - 1)$ lattice with free and cylindrical boundary conditions, with a single monomer  on the boundary,
the partition functions of the anisotropic dimer model}  can be {written} in terms of a partition function with twisted-boundary conditions $Z_{\alpha,\beta}$ with  $(\alpha,\beta) =(0,0)$. 
Based on these {considerations}, the exact asymptotic expansions of the free energy were {calculated}.

Let us denote a connected graph by {$G = G(V,E)$} where $V$ is the set of its vertices and $E$ is the set of its edges.
A  spanning tree $T$ is a subgraph of $G$ which has {$|V|- 1$} edges with at least one {edge  at} each vertex (we consider the case without loops). The number of edges attached to a vertex is its degree or coordination number.
In 2000, closed-form expressions for the spanning-tree generating function were derived by Tzeng and Wu  for a $d$-dimensional hypercubic lattice with free and periodic boundary conditions and for a combination of the two.
Analogous results were obtained for a simple quartic net embedded on two nonorientable surfaces, namely the Klein bottle and M$\ddot{o}$bius strip \cite{Tzeng}.
In 2015 Izmailian and Kenna  considered five different sets of boundary conditions  for the spanning tree on finite square lattices, and expressed the partition functions in terms of a principal partition function with twisted-boundary conditions.
In each case they also derived the exact asymptotic expansions of the logarithm of the partition function  \cite{Izmailian2015a}.
Izmailian, Kenna and Wu also derived the spanning-tree-generating function for cobweb and  fan networks \cite{Izmailian2014,Izmailian2015}.

In this paper we  complement the above studies by deriving expressions for the generating function of the spanning tree on cobweb and fan networks in terms of a partition function with twisted-boundary conditions $Z_{0,1/2}(z, {\cal M}, {\cal N})$. The significance of this result is that it verifies the applicability of the methods and algorithms developed in Ref.  \cite{Izmailian2002}. To demonstrate the broad range of usage of said algorithm to hitherto unsolved problems we furthermore derive the exact asymptotic expansions of the logarithm of the generating function for all networks mentioned above. In all cases we show that the exact asymptotic expansion of the free energy takes the form
\begin{equation}
f=f_{\text{bulk}}+\frac{2f_{1s}}{M}+\frac{2f_{2s}}{N}+f_{\text{corn}}\frac{\ln S}{S} + \frac{f_{0}(\rho)}{S}+ \sum_{p=1}^{\infty}\frac{f_{p}(\rho)}{S^{p+1}}\label{freeenergy}.
\end{equation}
Here $S=M \times N$ is the area of the lattice, $\rho = z \xi$ and $\xi=M/N$ is the aspect ratio. The term $z$ is
\begin{equation}
z=\frac{x}{y} \label{zx1x2}
\end{equation}
where $x$ and $y$ are the weights associated with the edges in the horizontal and vertical directions, respectively. The bulk free energy is $f_{\text{bulk}}$, the surface free energies are $f_{1s}$ and $f_{2s}$ and the corner free energy is $f_{\text{corn}}$.
The leading finite-size correction term  is  $f_0(\rho)$
and the subleading correction terms are  $f_p(\rho)$ for
$p = 1,2,3,...$.

The bulk free energy term $f_{\text{bulk}}$ is nonuniversal as are the surface free energies $f_{1s}$ and $f_{2s}$ and the subleading correction terms $f_p(\rho)$ $(p = 1,2,3, . . .)$.
In contrast, $f_{\text{corn}}$ is believed to be universal \cite{Privman,Cardy}. 
The leading finite-size correction term $f_0(\rho)$ is related to the conformal partition function and {in the 
limits ${\rho \to \infty}$ and ${\rho \to 0}$} its value is related to the conformal anomaly $c$ and conformal weights of the underlying conformal theory \cite{Blote,Affleck}. Moreover, in 1991, Kleban and Vassileva  \cite{Kleban} have shown that in {a} rectangular geometry on the plane the leading finite-size correction term  $f_0(\rho)$ contains  a geometry-dependent universal part  $f_\mathrm{univ}(\rho)$ given by
\begin{equation}
f_\mathrm{univ}(\rho)=\frac{c}{4}\ln{\left[\eta(\rho)\eta{(1/(\rho))}\right]}. \label{funiv}
\end{equation}
Here $\eta(\rho)$ is the Dedekind $\eta$ function.
However, Kleban and Vassileva  \cite{Kleban} mentioned that $f_0(\rho)$ can also contain  a non-universal additive constant $f_\mathrm{nonuniv}$,  which is not calculable via conformal field theory methods:
\begin{equation}
f_0(\rho)=f_\mathrm{univ}(\rho)+f_\mathrm{nonuniv}.\label{funnonun}
\end{equation}
{There} is little evidence {to support} these predictions from {either} exact solutions or numerical {determinations} \cite{Izmailian2014a,Izmailian2015a,Izmailian2012}. 
{An efficient bond propagation algorithm was recently used to compute} the partition function of the Ising model with free edges and corners in two dimensions on {a} rectangular lattice \cite{Izmailian2012}. 
{An efficient bond propagation algorithm was recently used to compute} the partition function of the Ising model with free edges and corners in two dimensions on {a} rectangular lattice \cite{Izmailian2012}. 
They verify the {predictions of } conformal field theory {presented in} by Eq. (\ref{funiv}) with central charge $c = 1/2$. 
Later the conformal field theory prediction Eq. (\ref{funiv}) was confirmed \cite{Izmailian2014a} for the dimer model on odd-odd square lattices with one monomer on the boundary, for which the central charge is $c=-2$ and for another model in the $c=-2$ universality class, {i.e.,} the spanning-tree model with free boundary conditions \cite{Izmailian2015a}. 
Moreover, the non-universal additive constant $f_\mathrm{nonuniv}$ has been determined in rectangular geometry in {the} Ising universality class \cite{Izmailian2012} and in $c=-2$ universality class for {the} dimer model  \cite{Izmailian2014a} and spanning tree model \cite{Izmailian2015a}. 
We are not aware of any other results similar to Eq. (\ref{funiv}) for other geometries, except the plane. 
For example, in the torus geometry, the leading finite-size correction term  $f_0(\rho)$ has been derived for three models on a torus {in  different} universality classes, namely the Ising ($c=1/2$) \cite{Ferdinand1}, dimer ($c=-2$) \cite{Ferdinand2} and Gaussian ($c=1$) models \cite{Izmailian2002}{. They take the form}
\begin{eqnarray}
f_0(\rho)&=& -\ln\frac{\theta_2+\theta_3+\theta_4}{2\eta(\rho)} \hspace{3.5 cm} \mbox{for {the} Ising model,}
\label{ising} \\
f_0(\rho)&=&-\ln\frac{\theta_2^2+\theta_3^2+\theta_4^2}{2\eta^2(\rho)} \hspace{3.5 cm} \mbox{for {the} dimer model,}
\label{dimer}\\
f_0(\rho)&=&\ln\sqrt{\rho}\,\eta^2(\rho)=\ln\eta(\rho)\eta(1/\rho) \hspace{1.8 cm}\mbox{for {the} Gaussian model}{.} \label{gauss}
\end{eqnarray}
One can see from Eqs. (\ref{ising}) - (\ref{gauss}) that in the torus geometry the leading finite-size correction term  $f_0(\rho)$ {cannot} be represented in {a form} similar to Eq. (\ref{funiv}). 
Nevertheless they are related to the conformal partition function in a torus geometry. 
For example, for the $c=1/2$ universality class the conformal partition function on torus ($Z$) is given by (see, for example, Ref. \cite{DiFranc}, page 349)
\begin{equation}
Z=\frac{\theta_2}{\eta }+\frac{\theta_3}{\eta}+\frac{\theta_4}{\eta}.\label{Zconf}
\end{equation}
It is easy to see from Eqs. (\ref{ising}) and (\ref{Zconf}) that for {the} Ising universality class in {the torus geometry,} the leading finite-size correction term  $f_0(\rho)$ is related to the conformal partition function on the torus $Z$. 
The same is true for other geometries and for other universality classes.

In this paper we  derive the leading finite-size correction terms $f_0(\rho)$ for the spanning tree model on cobweb and fan networks using the algorithm developed in Ref. \cite{Izmailian2002}. 
We  show to which conformal partition functions they are related.
We  also derive the leading finite-size correction term $f_0(\rho)$ for the Ising model on a cobweb network to see the difference between the corresponding conformal partition functions in the $c=-2$ and $c=1/2$ universality classes. We are also especially interested in the universal corner terms $f_{\text{corn}}$ because they are logarithmic. Using CFT, Cardy and Peschel predicted that a corner with an angle $\pi/2$ and two edges under free boundary conditions has
\begin{equation}
f_{\text{corn}}(0,0) = -\frac{c}{32}. \label{corner}
\end{equation}
In this formula, $c$ represents the central charge defining the universality class of the system \cite{Cardy}.
We confirmed this in Refs.\cite{Izmailian2012,Izmailian2013} for the square and triangular lattices with  free boundary conditions. Imamura et al. \cite{Imamura} and Bondesan et al. \cite{Bondesan2012,Bondesan2013} also used CFT to study the corner terms with different free boundary conditions   and found that the contribution to the free energy from a corner with two edges is
\begin{equation}
f_{\text{corn}}(\alpha \beta) = \Delta_{\alpha \beta} - \frac{c}{32}.
\label{fcornab}
\end{equation}
This formula,  where $\Delta_{\alpha \beta}$ represents the conformal weight of the boundary operator inserted at the corner, was verified in our previous work on Ising model on the square lattice  with different boundary conditions \cite{Izmailian2015c}.

\section{Spanning tree on networks}
\label{Spanning}
Let us consider the problem of enumerating weighted spanning trees on the $M \times N$ network.  The enumeration of spanning trees involves the evaluation of the tree generating function (or partition function) $Z_{\text{network}}^{\text{Sp}}$
\begin{equation}
Z_{\text{network}}^{\text{Sp}}({\cal L};x,y)=\sum_T x^{n_x}y^{n_y}
\label{span}
\end{equation}
where we assign weights $x$ and $y$, respectively, to edges in the horizontal and vertical directions. The summation is taken over all spanning tree configurations T on ${\cal L}$ and $n_x$ and $n_y$ are the numbers of edges in the spanning tree in the respective directions.

\begin{figure*}[tbp]
\includegraphics[width=0.9\textwidth]{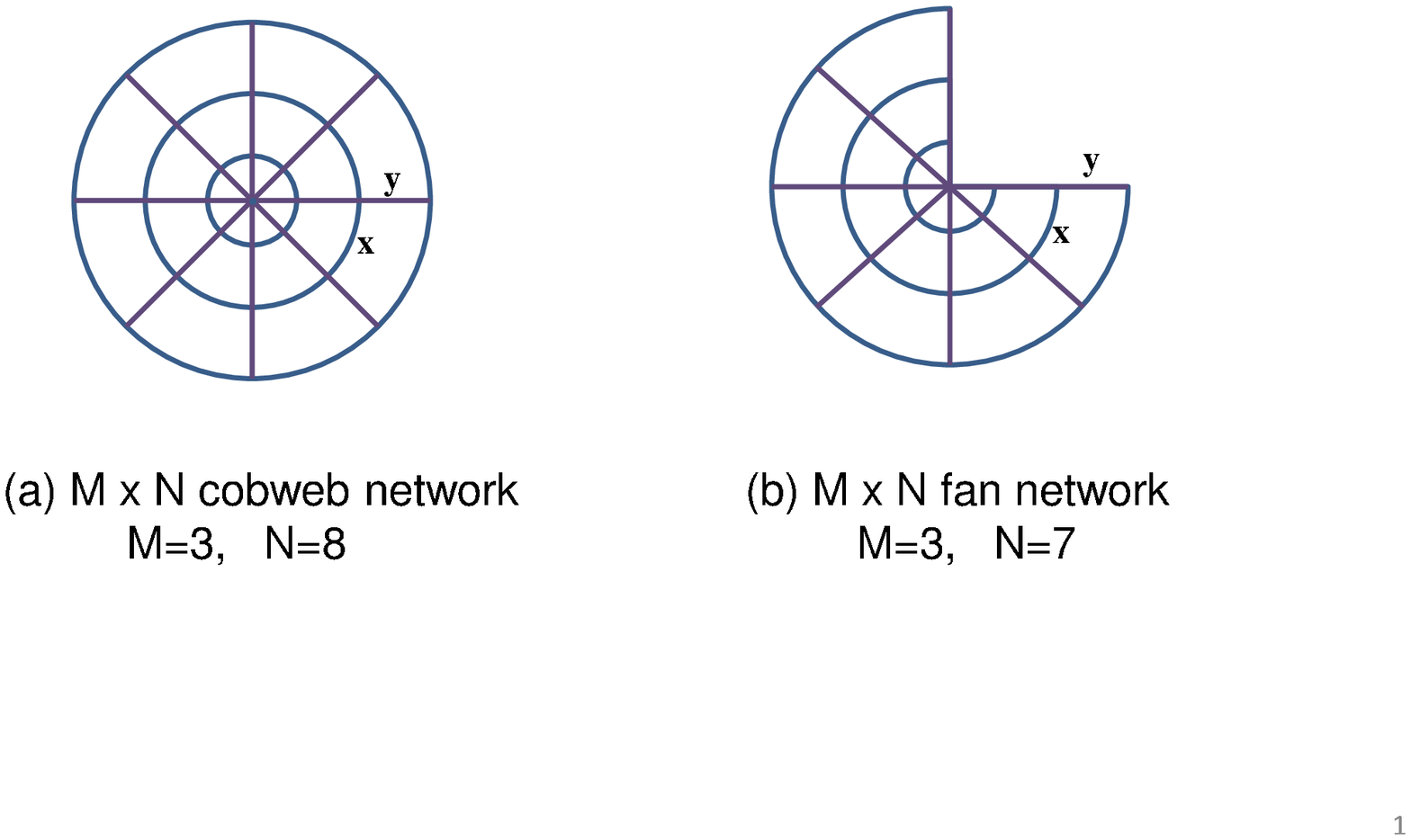}
	\caption{An $M \times N$ cobweb network with $M = 3$ and $N = 8$ (left). An $M \times N$ fan network with $M = 3$ and $N = 7$ (right). The weights $x$ and $y$ are assigned to the bonds in circular and radial directions, respectively. The cobweb network can be considered as a cylinder, where all sites on the one boundary are connected to an external common site, which is denoted by $0$,  while the fan network can be considered as a plane rectangular lattice, where all sites on the one of four boundaries are connected to an external common site.} \label{fig_1}
\end{figure*}

\subsection{Spanning tree on cobweb network}
\label{Spanning1}
The cobweb lattice ${\cal L}_{\text{cob}}$ is an $M \times N$ rectangular {lattice} with periodic boundary {conditions} in the horizontal direction and nodes on one of the two boundaries in the other direction connected
to an external common node. Therefore there is a total of $M N+1$ nodes and $2 M N$ edges. Topologically ${\cal L}_{\text{cob}}$ is of the form of a wheel consisting of $N$ spokes and $M$ concentric circles (see Fig. \ref{fig_1}) where the circumference of the circle corresponds to the ``horizontal" direction and spokes correspond to the ``vertical" direction. Note that the cobweb lattice ${\cal L}_{\text{cob}}$ can be considered as rectangular self-dual lattices~\cite{LuWu}.
The tree generating function (or partition function) for the
spanning tree model on cobweb network has been obtained in \cite{Izmailian2014} and can be written as
\begin{eqnarray}
Z_{\text{cobweb}}^{\text{Sp}}({\cal L};x,y)&=& y^{M
	N}\prod_{n=0}^{N-1}\prod_{m=0}^{M-1}4\left[z \sin^2{\frac{\pi
		n}{N}}+\sin^2{\frac{\pi (m+1/2)}{2M+1}}\right]
\label{spangenc3}
\end{eqnarray}
where $z=x/y$ and $x$ and $y$ are weights of the edges in the spoke and circle directions, respectively.

It has been shown \cite{Izmailian2002,Izmailian2003,Izmailian2006,Izmailian2015a} that the exact partition functions of the Ising model, dimer model and spanning tree model on different planar lattices under free, cylindrical and periodic boundary conditions can be {written} in terms of the single expression $Z_{\alpha,\beta}(z,{\cal M},{\cal N})$ with $(\alpha, \beta)= (1/2, 0),~(0,
1/2)$ and $(1/2, 1/2)$, where
\begin{equation}
Z^2_{\alpha,\beta}(z,{\cal M},{\cal N})=
\prod_{m=0}^{{\cal M}-1}\prod_{n=0}^{{\cal N}-1}4\left[z \sin^2\frac{(n+\alpha)\pi}{{\cal N}}+ \sin^2\frac{(m+\beta) \pi}{\cal M}\right], \label{Zab}
\end{equation}
with $(\alpha, \beta) \ne (0, 0)$. Note that the general theory about the asymptotic expansion of $Z_{\alpha,\beta}(z,{\cal M},{\cal N})$ has been given in \cite{Izmailian2002,Izmailian2003}.

In what follows, we will show that the tree generating function on
the cobweb network can be expressed in terms of  $Z_{0,1/2}(z,{\cal M},{\cal N})$ with ${\cal M} = 2M +1$ and ${\cal N} = N$
\begin{equation}
Z^2_{0,1/2}(z,{2M+1},{N})= \prod_{n=0}^{N-1}{\prod_{m=0}^{2M}}4\left[\textstyle{z\,\sin^2\frac{n \pi}{N}+
	\sin^2\frac{(m+1/2) \pi}{{2 M+1}}}\right]. \label{Z00}
\end{equation}
First we express double products $\prod_{n=0}^{N-1}
{\prod_{m=0}^{2M}} f(n,m)$ in terms of
$\prod_{n=0}^{N-1} \prod_{m=0}^{M-1}f(n,m)$, where
\begin{equation}
f(n,m)=4\left[\textstyle{z\,\sin^2\frac{n \pi}{N} + \sin^2\frac{(m+1/2) \pi}{2 M+1}}\right]. \label{transfc1}
\end{equation}
It is easy to show that
\begin{equation}
\prod_{n=0}^{N-1}{\prod_{m=0}^{2M}}
f(n,m)= \left({\prod_{n=0}^{N-1}f(n,M)}\right)
\left(\prod_{n=0}^{N-1}
\prod_{m=0}^{M-1}f(n,m)\right)^2, \label{transfc2}
\end{equation}
with $f(n,M)=4\left(z\,\sin^2\frac{n \pi}{N}+1\right)$.

With the help of the identity \cite{GradshteinRyzhik}
\begin{eqnarray}
\prod_{k=0}^{K-1}4\textstyle{ \left[~\!\sinh^2\omega +
	\sin^2\frac{k \pi }{K}\right]}&=& 4 \sinh^2\left(K\; \omega
\right),
\label{identity}
\end{eqnarray}
the product $\prod_{n=0}^{N-1}f(n,M)$ can be written as
\begin{eqnarray}
\prod_{n=0}^{N-1}f(n,M)&=&4 z^N \sinh^2{\left[
	N{\rm arcsinh}\, \sqrt{1/z} \right]}.
\label{prod}
\end{eqnarray}
Now using  Eqs. (\ref{spangenc3}) - (\ref{transfc2}),
(\ref{prod}) the tree generating function on the
cobweb network can be expressed finally as
\begin{eqnarray}
Z_{\text{cobweb}}^{\text{Sp}}({\cal L};x,y) =Q_1 \; Z_{0,1/2}(z,2M+1,N),
\label{Zfree20}
\end{eqnarray}
with
\begin{eqnarray}
Q_1 = \frac{y^{M N}}{2 z^{N/2} \sinh{\left[
		N{\rm arcsinh}\, \sqrt{1/z} \right]}}. \label{Zfree201}
\end{eqnarray}
Thus we have linked the cobweb partition function to the more general expression $Z_{\alpha,\beta}(z,{\cal M},{\cal N})$, further extending the applicability of the latter.

\subsection{Spanning tree on fan network}
\label{Spanning2}
The fan lattice ${\cal L}_{\text{fan}}$ is an $M \times N$ rectangular lattice of $M$ rows and $N$ columns with free boundary conditions on three sides of the lattice and nodes on the fourth boundary connected to an external additional node. Therefore there is a total of $M N+1$ nodes and $2 M N - M$ edges. We use the term Dirichlet-Neumann to describe the boundary conditions along the fourth boundary. Topologically, the obtained lattice is of the form of ''fan" consisting of $N$ radial lines and $M$ transverse arcs (see Fig. \ref{fig_1}), where transverse arcs correspond to the ''horizontal" direction and radial lines corresponds to the ''vertical" directions. In other words we impose Neumann or free boundary conditions along the two border spokes and along the outermost arc. We use the term Dirichlet-Neumann to describe the boundary conditions along the innermost arc.

The tree generating function for the
spanning tree model on {the} fan network has been obtained in \cite{Izmailian2014} and can be written as
\begin{eqnarray}
Z_{\text{fan}}^{\text{Sp}}({\cal L};x,y) &=& y^{M N}\prod_{m=0}^{M-1}\prod_{n=0}^{N-1}4\left[z\,\sin^2{\frac{\pi n}{2N}}+\sin^2{\frac{\pi (m+1/2)}{2M+1}}\right],
\label{spangen3}
\end{eqnarray}
where $z=x/y$.

In what follows, we will show that the tree generating function on the fan network can be expressed in terms of the single quantity
$Z_{0,1/2}(z,2M+1,2N)$
\begin{equation}
Z^2_{0,1/2}(z,2M+1,2N)=\prod_{n=0}^{2N-1}\prod_{m=0}^{2M}4\left[\textstyle{z\,\sin^2\frac{n \pi}{2N}+\sin^2\frac{(m+1/2) \pi}{2M+1}}\right].
	\label{Z120h}
\end{equation}
Now we first express double products $\prod_{n=0}^{2N-1}
{\prod_{m=0}^{2M}} f(n,m)$ in terms of
$\prod_{n=0}^{N-1} \prod_{m=0}^{M-1}f(n,m)$, where
\begin{equation}
	f(n,m)=4\left[\textstyle{z\,\sin^2\frac{n \pi}{2 N}+\sin^2\frac{(m+1/2) \pi}{2 M+1}}\right]. \label{transf1}
\end{equation}
It is easy to show that $f(2N-n,m)=f(n,2M-m)=f(n,m)$ and thus
\begin{equation}
	\prod_{n=0}^{2N-1}{\prod_{m=0}^{2M}}
	f(n,m)=\frac{\prod_{n=0}^{N-1}f(n,M)^2
		\prod_{m=0}^{2M}f(N,m)}{\prod_{m=0}^{2M}f(0,m)}\left(\prod_{n=0}^{N-1}
	\prod_{m=0}^{M-1}f(n,m)\right)^4. \label{transf2}
\end{equation}

With the help of the identities \cite{GradshteinRyzhik}
\begin{eqnarray}
	\prod_{k=0}^{K-1}4\textstyle{ \left[~\!\sinh^2\omega +
		\sin^2\frac{k \pi }{K}\right]}&=& 4 \sinh^2\left(K\; \omega
	\right),
	\label{identity1}\\
	\prod_{k=0}^{K-1}4\textstyle{ \left[~\!\sinh^2\omega +
		\sin^2\frac{(k+1/2) \pi }{K}\right]} &=& 4 \cosh^2\left(K\; \omega \right), \label{identity2}\\
	\prod_{m=0}^{M-1}4 \sin^2\frac{(m+1/2) \pi}{M}&=&4,
	\label{identity3}
\end{eqnarray}
the products $\prod_{m=0}^{2M}f(N,m)$, $\prod_{n=0}^{N-1}f(n,M)^2$
and $\prod_{m=0}^{2M}f(0,m)$ can be written as
\begin{eqnarray}
	\prod_{n=0}^{N-1}f(n,M)^2&=&\frac{4z^{2N}}{1+z} \sinh^2{\left[2
		N{\rm arcsinh}\,\sqrt{1/z} \right]}
	\label{prod1}\\
	\prod_{m=0}^{2M}f(N,m)&=&4 \cosh^2{\left[(2 M+1){\rm
			arcsinh}\, \sqrt{z}\right]}, \label{prod2}\\
	\prod_{m=0}^{2M}f(0,m)&=&4. \label{prod3}
\end{eqnarray}
Now using  Eqs. (\ref{spangen3}) - (\ref{transf2}),
(\ref{prod1}) - (\ref{prod3}) the tree generating function on the
fan network can finally be expressed   as
\begin{eqnarray}
	Z_{\text{fan}}^{\text{Sp}}({\cal L};x,y) =Q_2 \; Z^{1/2}_{0,1/2}(z,2M+1,2N),
	\label{Zfree2}
\end{eqnarray}
with
\begin{eqnarray}
	Q_2 = \frac{y^{M N}(1+z)^{1/4}}{\sqrt{2 z^N \sinh{\left[2 N{\rm arcsinh}\, \sqrt{1/z} \right]}\cosh{\left[(2 M+1){\rm arcsinh}\,\sqrt{z} \right]}}}, \label{Zfree21}
\end{eqnarray}
Again  we have  extended the applicability of the  general expression
$Z_{\alpha,\beta}(z,{\cal M},{\cal N})$  by linked it to the fan partition function.

\section{Asymptotic expansion of free energy}
\label{free energy}
Thus we have expressed the generating functions of the spanning tree on cobweb and fan networks in terms of a principal partition function with twisted-boundary conditions $Z_{0,1/2}(z,{\cal M},{\cal N})$ only. Based on such results, one can use the exact asymptotic expansions of $Z_{0,1/2}(z,{\cal M},{\cal N})$ given in Ref. \cite{Izmailian2002} to derive the exact asymptotic expansions of the free energy of the spanning tree $f = - \frac{1}{S} \ln Z$ for all networks mentioned above in terms of the Kronecker’s double series \cite{Izmailian2002}, which are directly related to elliptic $\theta$ functions. For the reader's convenience the asymptotic expansion of   $\ln Z_{0,1/2}(z,{\cal M},{\cal N})$ is given in Appendix~A.

Using Eq. (\ref{ExpansZab}) we can easily write down all the terms of the exact asymptotic expansion Eq. (\ref{freeenergy}) of the free energy,  $f = - \frac{1}{S} \ln Z$  for all models under consideration.

The bulk free energy $f_{\rm{bulk}}$ in Eq. (\ref{freeenergy}) for the weighted spanning tree on finite $M \times N +1$ lattices for all networks is given by
\begin{eqnarray}
f_{\rm{bulk}}&=&-\ln y-\frac{2}{\pi}\int_0^\pi\omega_z(k)dk =-\ln y-\frac{1}{\pi}\sum_{n=0}^{\infty}(-1)^n(n+1/2)^{-2}z^{n+1/2}\nonumber\\
&=&-\ln y-\frac{z^{1/2}\,\Phi(-z,2,\frac{1}{2})}{\pi},
\label{fbulksp}
\end{eqnarray}
where $\omega_z(k)$ is given by Eq. (\ref{SpectralFunction1}) and $\Phi(-z,2,1/2)$ is the Lerch transcendent. In particular, for isotropic spanning tree ($z=1$), the Lerch transcendent is now $\Phi(-1,2,{1}/{2})=4 G$, where $G = 0.915 965 594 \dots$ is the Catalan constant.

\subsection{Asymptotic expansion of free energy of the spanning tree on cobweb network}

\label{Expansion Z(1/2,0)(z,M,N)}

Using Eqs. (\ref{Zfree20}), (\ref{Zfree201}) and (\ref{ExpansZab}), the exact asymptotic expansions of the free energy for the spanning tree on the cobweb network, $f = - \frac{1}{S} \ln Z_{\text{cobweb}}^{\text{Sp}}$, can be written as
\begin{eqnarray}
f &=& - \frac{1}{S} \ln Z_{\text{cobweb}}^{\text{Sp}}=-\ln y +\frac{\frac{1}{2}\ln x+ {\rm arcsinh}\,\frac{1}{ \sqrt{z}}}{\left(M+\frac{1}{2}\right)}- \frac{1}{S}\ln Z_{0,1/2}(z,2M+1,N)\nonumber\\
&=& f_{\rm{bulk}}+\frac{2f_{1s}}{\left(M+\frac{1}{2}\right)}
- \frac{1}{S} \ln \frac{\theta_{2}(2\,\sqrt{z}\, \xi)}{\eta(2\, \sqrt{z}\, \xi)}+\frac{4\pi\xi}{S}\sum_{p=1}^\infty
\left(\frac{\pi^2 \xi}{S}\right)^p\frac{\Lambda_{2p}}{(2p)!}
\frac{K_{2p+2}^{\frac{1}{2},0}(2\,i\, \sqrt{z}\, \xi)}{2p+2}, \label{ExpansZcobweb}
\end{eqnarray}
where $f_{\rm{bulk}}$ is given by Eq. (\ref{fbulksp}), the surface free energy  $f_{2s}$ is equal to zero and  $f_{1s}$ is given by
\begin{equation}
f_{1s} =\frac{1}{4}\ln x+\frac{1}{2}{\rm arcsinh}\, \frac{1}{\sqrt{z}}, \label{f1sc}
\end{equation}
and $S$ and $\xi$ are given by
\begin{equation}
S = \left(M+\frac{1}{2}\right) N, \qquad \xi = \frac{M+\frac{1}{2}}{N}. \label{areac}
\end{equation}
The expression for $S$ is neither equal to the number of nodes or of edges; rather it emerges from the asymptotic expansion of the logarithm of $Z_{0,1/2}(z,2M+1,N)$. 
Thus the exact asymptotic expansions of the free energy for the spanning tree on the cobweb network can be written in the form given by Eq. (\ref{freeenergy}). For the leading correction terms $f_0(z\xi)$ we obtain
\begin{eqnarray}
f_0(z \xi)&=&-\ln \frac{\theta_{2}(2\sqrt{z}\,\xi)}{\eta( 2\sqrt{z}\, \xi)}.
\label{f0cobsp}
\end{eqnarray}
To check whether or not the leading finite-size correction term  $f_0(\rho)$ given by Eq. (\ref{f0cobsp}) can be {represented} in a form similar to Eq. (\ref{funiv}) we have to consider the Ising model ($c=1/2$) on the cobweb network \cite{Wu1982}. We have obtained that the exact asymptotic expansions of the free energy for the  Ising model on the cobweb network can be written in the form given by Eq. (\ref{freeenergy}) with $z=1$ (the details of the calculation will be reported elsewhere). The bulk free energy $f_{\rm{bulk}}$  for the Ising model on the cobweb network is given by
\begin{equation}
f_{\rm{bulk}}=-\frac{1}{2}\ln 2-\frac{2G}{\pi}.
\label{fbulkising}
\end{equation}
The surface free energy $f_{2s}$ is equal to zero and $f_{1s}$ is given by
\begin{equation}
f_{1s} =\frac{1}{2}\ln(1+\sqrt{2})-\frac{1}{8}\ln{2}-\frac{1}{4\pi}\int_0^{\pi}\ln{\left(\sqrt{2}\sin x+\sqrt{1+\sin^2 x}\right)}dx, \label{f1scising}
\end{equation}
and the leading finite-size correction term  $f_0(\xi)$ for the Ising model in the rectangular geometry on the cobweb network is given by
\begin{equation}
f_0(\xi)=-\frac{1}{2}\ln \frac{2\,\theta_{4}(2\,\xi)}{\eta( 2\, \xi)}.
\label{f0cobIsing}
\end{equation}
Note, that $S$ and $\xi$ for Ising model on the cobweb network are also given by Eq. (\ref{areac}).

Thus, we can see from Eqs. (\ref{f0cobsp}) and (\ref{f0cobIsing}) that the leading finite-size correction term  $f_0(\xi)$  cannot be represented in a form similar to Eq. (\ref{funiv}). Instead, based on Eq.  (\ref{f0cobsp}) and (\ref{f0cobIsing}), we can predict that the conformal partition function on the cobweb network for the Ising ($c=1/2$) universality class $Z_{\text{Ising}}$ is given by
\begin{equation}
Z_{\text{Ising}}=\sqrt{\frac{2\,\theta_{4}(2\,\xi)}{\eta( 2\, \xi)}}
\label{ZcobIsing}
\end{equation}
and for the spanning tree ($c=-2$) universality class the conformal partition function on cobweb network $Z_{\text{Sp}}^{\text{cobweb}}$ is given by
\begin{equation}
Z_{\text{Sp}}^{\text{cobweb}}=\frac{\theta_{2}(2\sqrt{z}\, \xi)}{\eta(2\sqrt{z}\, \xi)}
\label{Zcobsp}.
\end{equation}

For subleading correction terms $f_p(z \xi)$ for $p=1, 2, 3,\dots,$ we get
\begin{eqnarray}
f_p(z\, \xi)&=&\frac{2 \pi^{2p+1} \xi^{p+1}}{(2p)!(p+1)}\Lambda_{2p}\, K_{2p+2}^{0,1/2}(2 i \sqrt{z}\,\xi). \nonumber
\end{eqnarray}
The coefficients $\Lambda_{2p}$ are given in Eq. (\ref{omega}) and the Kronecker's double series
$K_{2p+2}^{0,1/2}(i z \xi)$ in terms of the elliptic theta functions are given in \cite{Izmailian2002,Izmailian2003} for arbitrary $p$.

It is easy to see from Eq. (\ref{ExpansZcobweb}), that the exact asymptotic expansions of the free energy for the spanning tree on finite cobweb network does not contain the corner free energy $f_{\rm{corner}}$, as it should be, since the logarithmic corner corrections to the free energy density should be absent for the systems without corners.

\subsection{Asymptotic expansion of free energy of the spanning tree on the fan network}

Using Eqs. (\ref{Zfree2}), (\ref{Zfree21}) and (\ref{ExpansZab}), the exact asymptotic expansions of the free energy for the spanning tree on the fan network, $f = - \frac{1}{S} \ln Z_{\text{fan}}^{\text{Sp}}$ can be written as
\begin{eqnarray}
f &=& - \frac{1}{S} \ln Z_{\text{fan}}^{\text{Sp}}\nonumber\\
&=&-\ln y +\frac{\frac{1}{2}\ln x+ {\rm arcsinh}\,\frac{1}{ \sqrt{z}}}{\left(M+\frac{1}{2}\right)}+\frac{{\rm arcsinh}\, \sqrt{z}}{N}- \frac{1}{4S}\ln 4(1+z)- \frac{1}{2S}\ln Z_{0,1/2}(z,2M+1,2N)\nonumber\\
&=& f_{\rm{bulk}}+\frac{2f_{1s}}{\left(M+\frac{1}{2}\right)}+\frac{2f_{2s}}{N}- \frac{1}{4S}\ln{4(1+z)}
- \frac{1}{2S} \ln \frac{\theta_{2}(\sqrt{z}\, \xi)}{\eta( \sqrt{z}\, \xi)}\nonumber\\
&+&\frac{\pi\xi}{S}\sum_{p=1}^\infty
\left(\frac{\pi^2 \xi}{4S}\right)^p\frac{\Lambda_{2p}}{(2p)!}
\frac{K_{2p+2}^{0,\frac{1}{2}}(i \sqrt{z}\, \xi)}{2p+2}, \label{ExpansZfan}
\end{eqnarray}
where $f_{\rm{bulk}}$ is given by Eq. (\ref{fbulksp}), the surface free energies  $f_{1s}$ is given by Eq. (\ref{f1sc}) and $f_{2s}$ is given by
\begin{equation}
 f_{2s} = \frac{1}{2}{\rm arcsinh}\, \sqrt{z}, \label{f2sf}
\end{equation}
and $S$ and $\xi$ are {again given by Eq.(\ref{areac}).} 
{The} exact asymptotic expansions of the free energy for the spanning tree on the fan network can {again} be written in the form given by Eq. (\ref{freeenergy}). 
For the leading correction terms $f_0(z\xi)$ we obtain
\begin{eqnarray}
f_0(z \xi)&=&-\frac{1}{4}\ln4(1+z)-\frac{1}{2}\ln \frac{\theta_{2}(\sqrt{z}\, \xi)}{\eta(\sqrt{z}\, \xi)}.
\label{f0fansp}
\end{eqnarray}
Thus, from Eq. (\ref{f0fansp}) one can see that a geometry-dependent universal part of the free energy $f_\mathrm{univ}(z\xi)$ in the rectangular geometry on the fan network is given by
\begin{equation}
f_\mathrm{univ}(z\xi)=-\frac{1}{2}\ln \frac{\theta_{2}(\sqrt{z}\, \xi)}{\eta(\sqrt{z}\, \xi)} \label{funivfan}
\end{equation}	
while a non-universal additive constant $f_\mathrm{nonuniv}$ is given by
\begin{equation}
f_\mathrm{nonuniv}=-\frac{1}{4}\ln4(1+z). \label{fnonunivfan}
\end{equation}	
As for the case of cobweb network we can predict that the conformal partition function for $c=-2$ universality class on the fan network $Z_{\text{Sp}}^{\text{fan}}$ is given by
\begin{equation}
Z_{Sp}^{\text{fan}}=\sqrt{\frac{\theta_{2}(\sqrt{z}\, \xi)}{\eta(\sqrt{z}\, \xi)}}
\label{funivfansp}
\end{equation}
It will be interesting to check  whether or not the leading finite-size correction term  $f_0(\sqrt{z}\, \xi)$ given by Eq. (\ref{f0fansp}) can be represented in a form similar to Eq. (\ref{funiv}) by considering models in different universality classes on {the} fan network, as well as to compute that term by conformal field theory method.

For subleading correction terms $f_p(z \xi)$ for $p=1, 2, 3,\dots, $ we get
\begin{eqnarray}
f_p(z\, \xi)&=&\frac{\pi^{2p+1} \xi^{p+1}}{4^p (2p)!(2p+2)}\Lambda_{2p}\,K_{2p+2}^{0,1/2}(i \sqrt{z}\, \xi).
\label{fpfannetwork}
\end{eqnarray}
The coefficients $\Lambda_{2p}$ are given in Eq. (\ref{omega}) and Kronecker's double series
$K_{2p+2}^{0,1/2}(i z \xi)$ in terms of the elliptic theta functions are given in \cite{Izmailian2002,Izmailian2003}.

Since the fan network is the plane rectangular lattice with three free boundary conditions and one with Dirichlet-Neumann boundary conditions, we have four corners for the fan network and one can expect the corner free energy $f_{\rm{corner}}$ contribution in the exact asymptotic expansions of the free energy Eq. (\ref{ExpansZfan}). But it is easy to see from Eq. (\ref{ExpansZfan}), that in the exact asymptotic expansions of the free energy for the spanning tree on the finite fan network the corner free energy $f_{\rm{corner}}$ is equal to zero.  Let us consider the corners of the fan network. Two of these corners have two edges each of which are subject to free boundary conditions and two corners have two edges of which one is under free and the other under Dirichlet-Neumann boundary conditions.  Thus the corner free energy $f_{\rm{corner}}$ can be written as a sum of four corner contribution, namely
\begin{equation}
f_{\rm{corner}}=2 f_{\text{corn}}(0,0)+2 f_{\text{corn}}(0,\beta) \label{fcorntotal}
\end{equation}
where $f_{\text{corn}}(0,0)$ is the contribution to the free energy from the corner with two edges with free boundary conditions, which is given by Eq. (\ref{corner})  with central charge $c$ equal to $c = - 2$, namely
$$
f_{\text{corn}}(0,0)=\frac{1}{16}
$$
and $f_{\text{corn}}(0,\beta)$ is the contribution to the free energy from the corner with two edges one under free and another under Dirichlet-Neumann boundary conditions,  which is given by Eq. (\ref{fcornab}) with central charge $c = - 2$ and $\Delta_{0,\beta} = -1/8$. Here   $\Delta_{0,\beta} = -1/8$ is the conformal weight of the boundary operator inserted at that corner. Thus $f_{\text{corn}}(0,\beta)$ is equal to
$$
f_{\text{corn}}(0,\beta)=\Delta_{0,\beta} - \frac{c}{32} = -\frac{1}{16}
$$
and the total contribution from the corners to free energy $f_{\text{corner}}$ is equal to zero. Although the fan network has four corners and each of these give corner contributions to the free energy, the sum of these contributions ($f_{\text{corner}}$) is equal to zero. Thus our result confirm both conformal theory \cite{Cardy,Imamura,Bondesan2012,Bondesan2013} and finite-size scaling \cite{privman1988} predictions.

\section{Spanning tree  on infinitely long strips}
\label{Spanninglong}
Finally, let us consider the case of an infinitely long strip. Conformal invariance implies that for an infinitely long two dimensional (2D) strip of finite width $L$ at criticality, the finite size scaling behavior of the critical free energy $f$ has the  form \cite{Blote,Affleck}
\begin{equation}
f=f_{\text{bulk}}+\frac{2f_{\text{surf}}}{L}+\frac{A}{L^2}+O\left(L^{-3}\right)\label{fconf}
\end{equation}
where the bulk free energy density $f_{\text{bulk}}$ and the surface free energy $f_{\text{surf}}$ are nonuniversal constants. In contrast, $A$ is a universal constants, but may depend on the boundary conditions. In some 2D geometries, the values of $A$  are known \cite{Blote,Affleck,Cardy1} to be related to the central charge $(c)$ and the conformal weight of the ground state $\Delta$
\begin{eqnarray}
A&=&4 \pi \zeta \left(\Delta-\frac{c}{24}\right) \qquad \qquad \mbox{in cylinder geometry}\label{cyl}\\
A&=& \pi \zeta \left(\Delta-\frac{c}{24}\right) \qquad \qquad \mbox{in strip geometry}\label{strip}
\end{eqnarray}
where the anisotropy factor $\zeta$ is a nonuniversal constant.

Let us consider the spanning tree case on the cobweb and fan network in the case when $M \to \infty$ (or $\xi \to \infty$). In that case the cobweb becomes an infinitely long cylinder with circumference $N$ and the fan network becomes an infinitely long strip of width $N$ and with free boundary condition on  both sides of the strip. The asymptotic expansion of the free energy for the cobweb and fan networks can be obtained from Eqs. (\ref{ExpansZcobweb}) and (\ref{ExpansZfan}), respectively. Using the facts that
\begin{eqnarray}
\lim_{\xi \to \infty}\theta_{2}(\sqrt{z}\, \xi) &=& \lim_{\xi \to \infty}2e^{-\frac{\pi \sqrt{z} \xi}{4}}=0 \label{thete2N}\\
\lim_{\xi \to \infty}\theta_{4}(\sqrt{z}\, \xi) &=& \lim_{\xi \to \infty}\theta_{3}(\sqrt{z}\, \xi)=1 \label{thete4N}\\
\lim_{\xi \to \infty}\eta(\sqrt{z}\, \xi) &=&\lim_{\xi \to \infty}e^{-\frac{\pi \sqrt{z} \xi}{12}}=0, \label{etaN}
\end{eqnarray}
the asymptotic expansion of the free energy for the cobweb  network on the infinitely long cylinder with circumference $N$ can be obtained from Eq. (\ref{ExpansZcobweb})
\begin{equation}
f=f_{\text{bulk}}+\frac{\pi \sqrt{z}}{3  N^2}+... .\label{fconfcyl}
\end{equation}
Thus by choosing $L=N$, the anisotropy factor $\zeta=\sqrt{z}$, the central charge $c=-2$ and conformal weight of the ground state $\Delta = 0$ we  get full agreement with conformal field predictions for cylinder geometry given by Eqs. (\ref{fconf}).

Again using Eqs. (\ref{thete2N}) and (\ref{etaN}) the asymptotic expansion of the free energy for the fan network on the infinitely long strip  with width $N$ and with free boundary condition on the both sides of the strip can be obtained from Eq. (\ref{ExpansZfan})
\begin{equation}
f=f_{\text{bulk}}+\frac{2f_{2s}}{N}+\frac{\pi  \sqrt{z}}{12 N^2}+\dots ,\label{fconfstripN}
\end{equation}
where $f_{2s}$ is given by Eq. (\ref{f2sf}). Thus by choosing $L=N$, the anisotropy factor $\zeta=\sqrt{z}$, the central charge $c=-2$ and conformal weight of the ground state $\Delta = 0$ we  get full agreement with conformal field predictions for strip geometry given by Eqs. (\ref{strip}).

Let us now consider the spanning tree case on the cobweb and fan networks in the case when $N \to \infty$ (or $\xi \to 0$). In that case both the cobweb and fan network become infinitely long strips with width $M$ and with free boundary condition on one side of the strip and  Dirichlet-Neumann boundary conditions on another side of the strip. The asymptotic expansion of the free energy for the cobweb and fan networks can be obtained from Eqs. (\ref{ExpansZcobweb}) and (\ref{ExpansZfan}), respectively.

To obtain the  asymptotic expansion of the free energy for cobweb and fan networks we need the behavior of the $\theta_{2}(\tau)$ - function and Dedekind’s $\eta(\tau)$ - function under the Jacobi transformation
$$
\tau \to \tau'=-1/\tau.
$$
The result for the $\theta_{2}(\tau)$-functions and Dedekind’s  $\eta(\tau)$
function is given in Appendex B of Ref. \cite{Korn}:
\begin{eqnarray}
\theta_{2}(\tau')&=&(- i \tau)^{1/2}\theta_4(\tau) \label{theta4yakob}\\
\eta(\tau')&=&(- i \tau)^{1/2}\eta(\tau) \label{etayakob}.
\end{eqnarray}
Using Eqs. (\ref{thete4N}), (\ref{etaN}), (\ref{theta4yakob}) and (\ref{etayakob}) one can obtain the asymptotic behavior of $\theta_2(\tau')$ and $\eta(\tau')$ as $\tau' \to 0$ (or $N \to \infty$).
Then from Eqs. (\ref{ExpansZcobweb}) and (\ref{ExpansZfan}) one can obtained the asymptotic expansion of the free energy for the cobweb and fan networks on the infinitely long strip in  the following form
\begin{equation}
f=f_{\text{bulk}}+\frac{2f_{1s}}{M+1/2}-\frac{\pi  }{24\sqrt{z}(M+1/2)^2}+...\label{fconfstrip}
\end{equation}
where $f_{1s}$ is given by Eq. (\ref{f1sc}). Thus by choosing $L=M+1/2$, the anisotropy factor $\zeta=1/\sqrt{z}$, the central charge $c=-2$ and conformal weight of the ground state $\Delta = -1/8$ we will get full agreement with conformal field predictions given by Eqs. (\ref{fconf}) and (\ref{strip}).

\section{Conclusions}
\label{Conclusion}
We analyzed spanning-tree generating functions for finite-size cobweb and fan networks and showed that each can be expressed in terms of a single, unifying partition function with twisted boundary conditions.
This reveals that the four corner free energies of the fan network cancel each other out so that their sum  matches the vanishing total value for the cobweb (which has no corners).

Thus we have extended the applicability of the twisted-boundary-method to a broad set of circumstances, opening up possible new approaches to efficiently investigate a multitude of spanning-tree problems of both fundamental and practical relevance.
For example, one could work backwards and seek different models which fit to the same $Z_{\alpha,\beta}$ and thus obey a class of ``strong'' universality.
To further demonstrate the power of the approach we have used it to derive exact finite-size corrections for the logarithm of the generating function of the spanning tree on both  these networks.
Then based the unified partition functions we  derived the exact asymptotic expansion of the logarithm of the partition function for the spanning tree on the cobweb and fan networks.
We also explain in the context of conformal field theory why the corner free energy for fan network, with its four corners, is equal to zero.
Based on our results for the leading finite-size correction term $f_0$ for the fan and cobweb networks we have predicted the conformal partition functions in $c=-2$ universality class for fan and cobweb geometries.
For the Ising model we have also predicted the conformal partition functions in {the} Ising ($c=1/2$) universality class for cobweb geometries.
And finally we have investigated the strip geometry for both above mentioned models and find an excellent agreement with conformal field theory predictions.
Thus we have confirmed universality for spanning trees and affirmed the robustness of the twisted-boundary-condition approach, opening new possible conduits to future research in a long-standing field of interest to a number of disciplines.

\section{Acknowledgment}
\label{Acknowledgment}

This work was partially supported by a grant of the Science Committee of the Ministry of Education and Science of the Republic of Armenia under contract 18T-1C113 and by the JINR program “Ter-Antonyan-Smorodinsky.”

\appendix

\section{Asymptotic expansion of and $Z_{0,\frac{1}{2}}(z,{\cal M},{\cal N})$}
\label{Expansion Z(0,0)(z,M,N)}

For the convenience of the reader, in this appendix we  present the exact asymptotic expansions of the logarithm of $Z_{0,\frac{1}{2}}(z,{\cal M},{\cal N})$ given in Ref. \cite{Izmailian2002}.

With the help of the identity
$$
4\left|~\!{\rm sinh}\left(M\omega+i\pi\beta\right)\right|^2 = 4\left[\,{\rm sinh}^2 M\omega + \sin^2\pi\beta\,\right]=\prod_{m=0}^{M-1}4\textstyle{ \left[~\!{\rm sinh}^2\omega + \sin^2\left(\frac{\pi (m+\beta)}{M}\right)\right]}
$$
the partition function with twisted boundary conditions
$Z_{\alpha,\beta}(z,{\cal M},{\cal N})$ given by Eq. (\ref{Zab}) can be transformed into {the} simpler form
\begin{equation}
Z_{\alpha,\beta}(z,{\cal M},{\cal N})=\prod_{n=0}^{N-1} 2\left| \textstyle{~\!{\rm sinh} \left[{\cal M}\omega_z\!\left(\frac{\pi(n+\alpha)}{\cal N}\right)+i\pi\beta \right] }\right| \label{Zabf}
\end{equation}
where lattice dispersion relation has been used
\begin{equation}
\omega_z(x)={\rm arcsinh}(\sqrt{z}\sin x).
\label{SpectralFunction1}
\end{equation}

The exact asymptotic expansion of the logarithm of $Z_{0,\frac{1}{2}}(z,{\cal M},{\cal N})$ in terms of the Kronecker's double series \cite{Izmailian2002,Weil} can be written as
\begin{eqnarray}
\ln Z_{0,\frac{1}{2}}(z,{\cal M},{\cal N})&=&\frac{{\cal S}}{\pi}\int_0^{\pi}
\omega_{z}(x)dx+
\ln \frac{\theta_{2}(\sqrt{z} \rho)}
{\eta(\sqrt{z} \rho)}-2\pi\rho\sum_{p=1}^\infty
\left(\frac{\pi^2 \rho}{{\cal S}}\right)^p\frac{\Lambda_{2p}}{(2p)!}
\frac{K_{2p+2}^{0,\frac{1}{2}}(i \sqrt{z} \rho)}{2p+2},
\label{ExpansZab}
\end{eqnarray}
where ${\cal S} = {\cal M} {\cal N}$, $\rho = {\cal M}/{\cal N}$, $\eta(\tau)$ is the Dedekind - $\eta$ function
\begin{equation}
\eta(\tau)=e^{\pi i \tau/12}\prod_{n=1}^{\infty}\left(1-e^{2 \pi i \tau n}\right). \label{eta}
\end{equation}
Dedekind's $\eta$ function satisfies the following identity:
\begin{equation}
2\eta(\tau)^3=\theta_2(\tau)\theta_3(\tau)\theta_4(\tau), \label{etaident}
\end{equation}
where $\theta_2(\tau), \theta_3(\tau), \theta_4(\tau)$ are elliptic theta functions. ${\rm K}_{2p}^{0,\frac{1}{2}}(\tau)$ is Kronecker's double series \cite{Izmailian2002,Weil}.

The differential operators $\Lambda_{2p}$ that  appear here can be expressed via coefficients $z_{2p}$ of the Taylor expansion of the lattice dispersion relation $\omega_{z}(x)$ 
\begin{equation}
\omega_{z}(x)=x\left(\sqrt{z} +\sum _{p=1}^{\infty} \frac{z
	_{2p}}{(2p)!}x^{2p}\right),
\label{SpectralFunction}
\end{equation}
	with $z_2=-\sqrt{z}(1+z)/3$, $z_4=\sqrt{z}(1+z)(1+9z)/5$,
$z_6=-\sqrt{z}(1+z)(1+90z+225z^2)/7$, etc.
\begin{eqnarray}
{\Lambda}_{2}&=&z_2,\nonumber\\
{\Lambda}_{4}&=&z_4+3z_2^2\,\frac{\partial}{\partial z},\nonumber\\
{\Lambda}_{6}&=&z_6+15z_4 z_2\,\frac{\partial}{\partial z}
+15z_2^3\,\frac{\partial^2}{\partial z^2}. \label{omega}\\
&\vdots&
\nonumber\\
{\Lambda}_{p}&=&\sum_{r=1}^{p}\sum
\left(\frac{ z_{p_1}}{p_1!}\right)^{k_1}\ldots
\left(\frac{ z_{p_r}}{p_r!}\right)^{k_r}\frac{p!}{k_1!\ldots
	k_r!}\;\frac{\partial^k}{\partial z^k}.
\label{L2p}
\end{eqnarray}
Here summation is over all positive numbers $\{k_1\ldots k_r\}$
and different positive numbers $\{p_1,\ldots,p_r\}$ such that $p_1
k_1+\ldots+ p_r k_r=p$ and $k=k_1+\ldots+k_r-1$.

The $\int_{0}^{\pi}\!\!\omega_z(x)~\!{\rm d}x$ is given by
\begin{equation}
\int_{0}^{\pi}\!\!\omega_z(x)~\!{\rm d}x =
\frac{z^{1/2}\,\Phi(-z,2,\frac{1}{2})}{2},
\end{equation}
where $\Phi(x,s,\alpha)$ is the Lerch transcendent defined as
\begin{equation}
\Phi(x,s,\alpha)=\sum_{n=0}^{\infty}(\alpha+n)^{-s}x^n.
\label{lerch}
\end{equation}
In particular, for the isotropic spanning tree ($z=1$), the Lerch transcendent is now
$\Phi(-1,2,{1}/{2})=4 G$, where $G$ is the Catalan constant given by
\begin{equation}
G=\sum_{n=0}^{\infty}{\frac{(-1)^n}{(2n+1)^2}}=0.915 965 594 \dots.
\label{Catelan}
\end{equation}
Kronecker's double series ${\rm K}_{2p}^{0,\frac{1}{2}}(\tau)$ can all be expressed in terms of the elliptic $\theta$-functions only \cite{Izmailian2002}.

\vspace{1cm}

{\Huge{Bibliography}}

\end{document}